# El Tutor de IA en la Formación de Ingenieros: Diseño, Resultados y Rediseño de una Experiencia de en Hidrología en una Universidad Argentina

*Hugo Roger Paz[1]*

hpaz**@herrrera.unt.edu.ar**

https://orcid.org/0000-0003-1237-7983

https://www.researchgate.net/profile/Hugo-Paz-3

## Resumen

La irrupción de la Inteligencia Artificial Generativa (IAGen) ha reconfigurado la educación superior, presentando oportunidades y desafíos ético-pedagógicos. Este artículo presenta un estudio de caso empírico sobre el ciclo completo (diseño, fracaso inicial, rediseño y re-evaluación) de una intervención con un Tutor de IA (ChatGPT) en la asignatura "Hidrología y Obras Hidráulicas" (Ingeniería Civil, UTN-FRT, Argentina).

El estudio documenta dos intervenciones en la misma cohorte (n=23). La primera resultó en un fracaso generalizado (0% de aprobados) por uso superficial y problemas graves de integridad académica (65% similitud, copias > 80%). Este fracaso forzó un rediseño metodológico integral. La segunda intervención, basada en un prompt rediseñado (Prompt V2) con controles estrictos de evidencia (Anexo A obligatorio con chat exportado, tiempo mínimo ≥ 120 minutos, ejercicio numérico verificable) y una rúbrica refinada (Rúbrica V2), mostró resultados significativamente mejores: mediana de 88/100 y cumplimiento verificable de procesos de interacción genuina.

Mediante un enfoque de métodos mixtos (análisis documental reproducible y análisis de rúbrica), se evalúa el impacto del rediseño en la integridad y el desempeño técnico. Los resultados demuestran que, sin controles de proceso explícitos, los estudiantes priorizan la eficiencia sobre el aprendizaje profundo, entregando documentos sin trazabilidad real. Se propone un protocolo de evaluación transferible para asignaturas STEM, centrado en "zonas personales auditables", para fomentar el pensamiento de orden superior. El estudio aporta evidencia empírica clave desde el contexto de una universidad pública latinoamericana.

**Palabras Clave:** Inteligencia Artificial Generativa, Educación en Ingeniería, Tutor Inteligente, Diseño Pedagógico, Evaluación del Aprendizaje.

# The AI Tutor in Engineering Education: Design, Results, and Redesign of an Experience in Hydrology at an Argentine University

## Abstract

The emergence of Generative Artificial Intelligence (GenAI) has reshaped higher education, presenting both opportunities and ethical-pedagogical challenges. This article presents an empirical case study on the complete cycle (design, initial failure, redesign, and re-evaluation) of an intervention using an AI Tutor (ChatGPT) in the "Hydrology and Hydraulic Works" course (Civil Engineering, UTN-FRT, Argentina).

The study documents two interventions in the same cohort (n=23). The first resulted in widespread failure (0% pass rate) due to superficial use and serious academic integrity issues (65% similarity, copies > 80%). This failure forced a comprehensive methodological redesign. The second intervention, based on a redesigned prompt (Prompt V2) with strict evidence controls (mandatory Appendix A with exported chat, minimum time ≥ 120 minutes, verifiable numerical exercise) and a refined rubric (Rubric V2), showed significantly better results: a median score of 88/100 and verifiable compliance with genuine interaction processes.

Using a mixed-methods approach (reproducible document analysis and rubric analysis), the impact of the redesign on integrity and technical performance is evaluated. The results demonstrate that, without explicit process controls, students prioritize efficiency over deep learning, submitting documents without real traceability. A transferable assessment protocol for STEM courses is proposed, centered on "auditable personal zones," to foster higher-order thinking. The study provides key empirical evidence from the context of a public Latin American university.

## O Tutor de IA na Formação de Engenheiros: Desenho, Resultados e Redesenho de uma Experiência em Hidrologia em uma Universidade Argentina

## Resumo

A emergência da Inteligência Artificial Generativa (IAGen) reconfigurou o ensino superior, apresentando oportunidades e desafios ético-pedagógicos. Este artigo apresenta um estudo de caso empírico sobre o ciclo completo (desenho, fracasso inicial, redesenho e reavaliação) de uma intervenção com um Tutor de IA (ChatGPT) na disciplina "Hidrologia e Obras Hidráulicas" (Engenharia Civil, UTN-FRT, Argentina).

O estudo documenta duas intervenções na mesma coorte (n=23). A primeira resultou em um fracasso generalizado (0% de aprovados) por uso superficial e problemas graves de integridade acadêmica (65% de similaridade, cópias > 80%). Esse fracasso forçou um redesenho metodológico integral. A segunda intervenção, baseada em um prompt redesenhado (Prompt V2) com controles rigorosos de evidência (Anexo A obrigatório com chat exportado, tempo mínimo ≥ 120 minutos, exercício numérico verificável) e uma rubrica refinada (Rubrica V2), mostrou resultados significativamente melhores: mediana de 88/100 e cumprimento verificável de processos de interação genuína.

Através de uma abordagem de métodos mistos (análise documental reproduzível e análise de rubrica), avalia-se o impacto do redesenho na integridade e no desempenho técnico. Os resultados demonstram que, sem controles de processo explícitos, os estudantes priorizam a eficiência em detrimento da aprendizagem profunda, entregando documentos sem rastreabilidade real. Propõe-se um protocolo de avaliação transferível para disciplinas STEM, focado em "zonas pessoais auditáveis", para fomentar o pensamento de ordem superior. O estudo fornece evidências empíricas cruciais do contexto de uma universidade pública latino-americana.

**Palavras-chave:** Inteligência Artificial Generativa, Educação em Engenharia, Tutor Inteligente, Desenho Pedagógico, Avaliação da Aprendizagem,.

## Le Tuteur IA dans la Formation des Ingénieurs : Conception, Résultats et Reconception d'une Expérience en Hydrologie dans une Université Argentine

**Résumé**

L'émergence de l'Intelligence Artificielle Générative (IAGen) a reconfiguré l'enseignement supérieur, présentant des opportunités et des défis éthiques et pédagogiques. Cet article présente une étude de cas empirique sur le cycle complet (conception, échec initial, reconception et réévaluation) d'une intervention avec un Tuteur IA (ChatGPT) dans le cours "Hydrologie et Ouvrages Hydrauliques" (Génie Civil, UTN-FRT, Argentine).

L'étude documente deux interventions au sein de la même cohorte (n=23). La première s'est soldée par un échec généralisé (0% de réussite) dû à une utilisation superficielle et à de graves problèmes d'intégrité académique (65% de similarité, copies > 80%). Cet échec a nécessité une reconception méthodologique complète. La seconde intervention, basée sur un prompt remanié (Prompt V2) avec des contrôles de preuve stricts (Annexe A obligatoire avec chat exporté, temps minimum ≥ 120 minutes, exercice numérique vérifiable) et une grille d'évaluation affinée (Grille V2), a montré des résultats significativement meilleurs : une médiane de 88/100 et un respect vérifiable des processus d'interaction authentique.

Grâce à une approche de méthodes mixtes (analyse documentaire reproductible et analyse de grille), l'impact de la reconception sur l'intégrité et la performance technique est évalué. Les résultats démontrent que, sans contrôles de processus explicites, les étudiants privilégient l'efficacité à l'apprentissage en profondeur, soumettant des documents sans traçabilité réelle. Un protocole d'évaluation transférable pour les cours STEM est proposé, centré sur des "zones personnelles auditables", pour encourager la pensée d'ordre supérieur. L'étude fournit des preuves empiriques clés issues du contexte d'une université publique latino-américaine.

**Mots-clés :** Intelligence Artificielle Générative, Formation des Ingénieurs, Tuteur Intelligent, Conception Pédagogique, Évaluation de l'Apprentissage.

# 1. Introducción: La Inteligencia Artificial Generativa en la Encrucijada de la Educación Superior Latinoamericana

El presente trabajo se inscribe en el marco de la convocatoria del dossier temático "La inteligencia artificial generativa (IAGen) en la docencia e investigación: buenas prácticas, retos y prospectiva" de la Revista Educación Superior y Sociedad (ESS), publicada por el Instituto Internacional de la UNESCO para la Educación Superior en América Latina y el Caribe (UNESCO IESALC). Desde el lanzamiento de ChatGPT por OpenAI en noviembre de 2022, la IAGen ha emergido como una fuerza disruptiva en la educación superior, posicionándose como un potencial "aliado para potenciar el conocimiento humano". No obstante, su rápida y masiva adopción ha generado una profunda reflexión sobre los "desafíos únicos que requieren una exploración detallada", especialmente en el contexto de América Latina.

La literatura internacional ha delineado rápidamente una dicotomía fundamental. Por un lado, existe un consenso creciente sobre el potencial transformador de estas herramientas para personalizar el aprendizaje, ofrecer retroalimentación inmediata y adaptativa, y automatizar tareas docentes, liberando tiempo para interacciones más significativas (Sabzalieva & Valentini, 2023; Arias et al., 2023). La IAGen promete sistemas de tutoría inteligente capaces de ajustarse al ritmo y estilo de cada estudiante, mejorando la experiencia educativa y los resultados del aprendizaje (Baidoo-Anu & Owusu Ansah, 2023). Por otro lado, esta promesa se ve contrapesada por desafíos éticos y prácticos de gran envergadura. La preocupción por la integridad académica se ha convertido en un tema central, con un aumento del riesgo de fraude y plagio que amenaza los cimientos de la evaluación tradicional (Gallent-Torres et al., 2023). A esto se suman inquietudes sobre la privacidad y seguridad de los datos de los estudiantes y el peligro de que la tecnología, en lugar de cerrar brechas, las amplíe, exacerbando las desigualdades existentes en la región.

Este panorama global adquiere matices particulares al anclarse en la realidad latinoamericana y, más específicamente, en Argentina. La adopción de la IAGen en la región no es un fenómeno homogéneo. Estudios comparativos de estudiantes de posgrado en Argentina, Brasil, Chile, Perú y España muestran una aceptación predominantemente positiva, aunque con reservas y preocupaciones éticas transversales a todos los países. En el contexto argentino, encuestas sobre los usos de ChatGPT revelan un dato de suma relevancia: el 85% de los usuarios que emplean la herramienta en tareas laborales perciben "ganancias de productividad horaria" (Zukerfeld et al., 2023). Este hallazgo sugiere que el principal motor de adopción de la IAGen en el país está fuertemente ligado a la eficiencia y la optimización del tiempo.

Esta orientación hacia la productividad genera una tensión fundamental con los objetivos pedagógicos de la educación superior. Mientras que los estudiantes, en un entorno cada vez más demandante, pueden percibir la IAGen como una herramienta para "hackear" tareas académicas y reducir el tiempo de dedicación, como lo demuestran casos de estudio internacionales donde los alumnos admiten resolver hasta el 80% de sus trabajos con IA, el rol de la universidad es fomentar el pensamiento crítico, la resolución de problemas complejos y el aprendizaje profundo. Esta divergencia entre la búsqueda de eficiencia por parte del estudiante y la búsqueda de profundidad por parte del educador constituye un desafío central.

Un diseño pedagógico que simplemente introduce la herramienta sin reestructurar la naturaleza de las tareas y la evaluación corre el riesgo de fomentar un aprendizaje superficial y una dependencia tecnológica que atenta contra el desarrollo de competencias clave. Por lo tanto, la integración de la IAGen debe ser diseñada de tal manera que su uso para el aprendizaje significativo sea más ventajoso y gratificante que su mero uso instrumental para completar una tarea.

A pesar del creciente volumen de revisiones sistemáticas y artículos teóricos, existe una notoria escasez de estudios de caso empíricos que documenten y analicen de manera rigurosa la aplicación de la IAGen en contextos específicos, particularmente en disciplinas técnicas como la ingeniería y en el ámbito de las universidades públicas de América Latina. Para contribuir a cerrar esta brecha, el presente artículo tiene como objetivo principal presentar un análisis exhaustivo del ciclo de vida completo — diseño, implementación, fracaso inicial, rediseño y re-evaluación exitosa— de una intervención pedagógica con un Tutor de IA en la asignatura "Hidrología y Obras Hidráulicas" de la Universidad Tecnológica Nacional - Facultad Regional Tucumán (UTN-FRT). A través de este estudio, se busca extraer buenas prácticas, identificar los retos inherentes a su aplicación en un contexto real y ofrecer una prospectiva informada que pueda guiar a otros educadores e instituciones de la región.

## 2. Marco Teórico

### 2.1 Inteligencia Artificial Generativa en la Educación Superior e Ingeniería

La Inteligencia Artificial Generativa, y en particular los modelos de lenguaje grandes (LLMs) como ChatGPT, han sido adoptados en educación por su capacidad de generar explicaciones, ejemplos numéricos y retroalimentación inmediata. En asignaturas de ingeniería, donde la solución de problemas numéricos y el razonamiento técnico son centrales, la IAGen puede actuar como tutor de práctica, simulador de diálogo experto y generador de plantillas de informe. La promesa es que estos sistemas pueden ajustarse al ritmo individual del estudiante, ofreciendo una experiencia de aprendizaje más personalizada y efectiva que las modalidades tradicionales de enseñanza masiva (Baidoo-Anu & Owusu Ansah, 2023; Tan et al., 2023).

Sin embargo, la literatura advierte sobre riesgos significativos: homogenización de respuestas, dependencia del estudiante que inhibe el desarrollo del pensamiento crítico, y dificultades para discernir la autoría cuando las salidas generadas por la IA se incorporan al trabajo final sin evidenciar el proceso de construcción del conocimiento. La implementación responsable requiere, por tanto, tres elementos fundamentales: (a) transparencia en el uso de la herramienta, (b) mecanismos de evidencia que permitan auditar el proceso de interacción (como la exportación de chats con marcas temporales), y (c) criterios evaluativos que distingan de manera clara la comprensión propia del estudiante de la mera repetición o presentación de salidas generadas por la IA (Aparicio-Gómez, 2023; Silgado-Tuñón & López-Flores, 2025).

En este contexto, la IAGen no debe ser conceptualizada como un mero repositorio de información o un "oráculo" al que se consulta para obtener respuestas directas, sino como un "compañero cognitivo" (cognitive partner) que sirve como andamiaje interactivo para el desarrollo del pensamiento de orden superior. Esta reconfiguración del rol exige un diseño instruccional que posicione la tecnología al servicio del aprendizaje profundo, y no como un atajo que lo elude.

## 2.2 Evaluación Asistida por IA e Integridad Académica

El uso de IAGen en evaluación abre dos frentes simultáneos y en tensión: por un lado, augura mayor escala y rapidez en la retroalimentación (una ventaja para sistemas educativos masificados); por otro, origina problemas críticos de atribución de autoría y riesgo de plagio asistido por IA. La disponibilidad de herramientas que pueden generar soluciones técnicas correctas y textos bien estructurados devalúa las formas de evaluación que se centran exclusivamente en el producto final, sin considerar el proceso mediante el cual el estudiante llegó a esa respuesta (García Huertes et al., 2024; Flanagan & Ogilvie, 2024).

Estudios recientes en el campo de la evaluación educativa recomiendan una transición decidida desde la evaluación del "producto" hacia la evaluación del "proceso". Este cambio paradigmático implica combinar: (i) control de procesos mediante el registro auditable de sesiones (timestampsstamps, logs de interacción), (ii) tareas que requieren síntesis personal y justificación explícita como prueba de comprensión, y (iii) análisis automatizado de similitud textual que incorpore no sólo medidas vectoriales clásicas como TF-IDF (Term Frequency-Inverse Document Frequency), sino también técnicas robustas ante parafraseo y reordenamientos como MinHash y Locality-Sensitive Hashing (LSH) para detectar coincidencias aproximadas en fragmentos largos (Zeng & Hovy, 2023; McCabe & Pavela, 2004).

La tensión entre la búsqueda de eficiencia del estudiante (motivado por las "ganancias de productividad" que la IA efectivamente ofrece) y la búsqueda de profundidad del educador (comprometido con el desarrollo de competencias de orden superior) no puede resolverse mediante prohibiciones o detección punitiva, sino mediante el diseño instruccional. La integración de la IAGen debe diseñarse de tal manera que el uso para el aprendizaje significativo sea estructuralmente más ventajoso, recompensado y verificable que el mero uso instrumental para completar una tarea sin comprensión.

## 3. Metodología

Para evaluar de manera integral la experiencia, se adoptó un diseño de estudio de caso empírico con un enfoque de métodos mixtos (mixed-methods approach), que permite combinar la profundidad del análisis cualitativo con la capacidad de generalización de los datos cuantitativos para obtener una comprensión holística del fenómeno. El estudio se estructuró en dos fases evaluativas distintas aplicadas a la misma cohorte de estudiantes.

### 3.1 Contexto y Participantes

La intervención se llevó a cabo en la asignatura "Hidrología y Obras Hidráulicas", una materia troncal del plan de estudios de Ingeniería Civil en la Universidad Tecnológica Nacional - Facultad Regional Tucumán (UTN-FRT). Esta asignatura se caracteriza por la necesidad de que los estudiantes integren conceptos teóricos abstractos, como el ciclo hidrológico y la estocástica de eventos extremos, con la aplicación de modelos matemáticos complejos para el diseño de infraestructura hidráulica.

Los participantes fueron 23 estudiantes de una cohorte regular del curso. Los desafíos pedagógicos tradicionales en esta asignatura incluyen la dificultad de los estudiantes para visualizar fenómenos dinámicos y la aplicación de fórmulas a problemas de ingeniería no estructurados del mundo real. Se garantizó el anonimato y la confidencialidad de todos los participantes y sus datos, de acuerdo con las normativas éticas de investigación vigentes.

### 3.2 Fase 1: Intervención Inicial (Informe 1 - Temas 1-3)

Objetivo: Evaluar la comprensión de los fundamentos teóricos (Ciclo Hidrológico, Precipitaciones, Estadística Hidrológica).

**Diseño del Tutor de IA (Prompt V1):** Se configuró un prompt maestro que establecía el rol del tutor (profesor/tutor experto en hidrología), la estructura modular de la interacción (cobertura de los tres temas fundamentales), y los criterios de salida (generación automática de un "Informe de la Sesión" técnico-académico al finalizar). El tutor debía explicar conceptos, realizar ejemplos ilustrativos y plantear preguntas de control para verificar la comprensión.

**Rúbrica de Evaluación V1:** La evaluación inicial incluía cuatro dimensiones con pesos diferenciados: Interacción (20%), Estructura (20%), Desempeño Técnico (35%) y Originalidad (25%). Se estableció un requisito de tiempo mínimo de interacción (≥ 60 minutos) y un umbral de similitud textual (< 75%) para la dimensión de originalidad. Sin embargo, la evidencia de interacción (exportación del chat completo) no era un requisito eliminatorio explícito en el prompt inicial, lo que constituiría una debilidad crítica del diseño.

**Requisitos de entrega:** Los estudiantes debían entregar un documento PDF que incluyera el informe generado por la IA y, opcionalmente, sus propias anotaciones o síntesis. La exportación del chat (transcripción completa de la interacción) era sugerida pero no obligatoria.

### 3.3 Resultados de la Fase 1 y Rediseño Metodológico

El análisis de la primera entrega reveló un fracaso metodológico y pedagógico generalizado, que motivó un proceso de auditoría exhaustiva y rediseño integral del procedimiento. Este análisis se realizó en cinco lotes sucesivos, cada uno revelando nuevas problemáticas que forzaron ajustes progresivos a los criterios de evaluación.

**Hallazgos principales del fracaso inicial:**

- **Falta de evidencia de proceso:** Solo 1 de 23 estudiantes (4%) adjuntó voluntariamente la transcripción completa del chat con el tutor. La ausencia de trazabilidad imposibilitó verificar si hubo interacción genuina o si los informes fueron generados retrospectivamente mediante prompts instrumentales.

- **Uso de plantillas y placeholders:** Se detectaron frases genéricas tipo "[aquí iría el tiempo]" o "[completar conclusiones]" en múltiples informes, indicando que los estudiantes habían solicitado a la IA la generación de una plantilla de informe sin completarla con contenido propio.

- **Falsa documentación del tiempo:** Varios estudiantes declararon bloques genéricos de "1 hora" o "2 horas" sin sellos horarios verificables. El único chat exportado mostró apenas 17 minutos reales de interacción, muy por debajo del mínimo esperado.

- **Ausencia de ejercicios numéricos:** Apenas 3 informes (13%) desarrollaron un cálculo aplicado (distribuciones de Gumbel, Log-Pearson III o curvas IDF). La falta de aplicación práctica contradecía directamente la finalidad pedagógica del tutor.

- **Copias no declaradas:** El análisis de similitud detectó dos pares de informes que superaron el 80% de similitud léxica, constituyendo casos claros de copia entre estudiantes. Otros nueve informes compartieron más del 55% de similitud por uso de la misma plantilla base generada por la IA.
- **Texto embebido en imágenes:** Un informe presentó el contenido como capturas de pantalla (texto renderizado como imagen), requiriendo extracción OCR. Aunque esto permitió rescatar el contenido, el caso reprobó igualmente por falta de evidencia de proceso.

**Resultado cuantitativo final:** Tras aplicar criterios de integridad estrictos en cinco rondas sucesivas de revisión, 0 de 21 informes válidos (0%) alcanzaron calificación aprobatoria. La nota media fue de 4.86 sobre 10. El 65% de los informes presentaron similitud ≥ 45% (umbral "medio" o superior), y múltiples casos fueron invalidados por falta completa de evidencia de interacción.

La conclusión factual fue contundente: la entrega masiva de informes sin trazabilidad real ni originalidad técnica desembocó en cero aprobaciones cuando se aplicaron criterios de integridad académica exigentes. Este fracaso evidenció que, sin controles de proceso explícitos, la IAGen facilitaba el plagio y el aprendizaje superficial, en lugar de promover la comprensión profunda.

### 3.4 Fase 2: Rediseño e Intervención (Informe 2 - Lluvia de Diseño)

El fracaso de la Fase 1 motivó un rediseño integral del procedimiento de evaluación, sistematizado en el documento interno "Informe de revisión y rediseño del procedimiento de evaluación con Tutor IA". El rediseño se basó en la premisa de que el diseño del prompt y la rúbrica es tan importante como la potencia técnica del Tutor de IA, y que sin requisitos estrictos de trazabilidad y originalidad, la mayoría de los estudiantes entregará documentos generados artificialmente sin aprendizaje real.

**Objetivo de la Fase 2:** Evaluar la aplicación práctica de conceptos avanzados (Lluvia de Diseño, Precipitación Efectiva, métodos de abstracciones como SCS-CN) mediante una interacción guiada que forzara la producción auditable del estudiante.

**Prompt V2 (Rediseñado):** El nuevo prompt fue significativamente más estricto y se estructuró en cinco módulos progresivos obligatorios (M1: Intensidad-Duración-Recurrencia local; M2: Distribución temporal de la lluvia; M3: Distribución areal; M4: Abstracciones/Pérdidas por método SCS-CN; M5: Integración final con ejemplo numérico). Cada módulo incluía checkpoints de comprensión, y el tutor debía guiar la resolución paso a paso de un ejemplo numérico estándar con una cuenca base de 10 km².

**Nuevos requisitos (claves del rediseño):**

- **"Anexo A" Obligatorio:** El estudiante debía exportar la transcripción completa del chat (con marcas temporales) y adjuntarla como Anexo A. La omisión de este anexo invalidaba automáticamente la actividad, sin importar la calidad aparente del informe.
- **Tiempo Mínimo Estricto:** El tiempo total de interacción, verificable en las marcas temporales del chat exportado, debía ser superior a 120 minutos. Tiempos inferiores invalidaban la actividad.
- **Ejemplo Numérico Base Verificable:** El tutor debía guiar un ejemplo numérico ilustrativo con cuenca de 10 km², y el estudiante debía demostrar comprensión mediante cálculos propios con unidades y justificación de parámetros.
- **Informe Final IA con Preguntas de Repaso:** Al finalizar la interacción, el tutor debía generar un "Informe Académico Final" que incluyera: resumen de temas cubiertos, tiempo total registrado, evaluación del desempeño del estudiante, cinco preguntas de repaso personalizadas (que el estudiante debía responder), sugerencias bibliográficas y una auto-evaluación de rúbrica.

**Rúbrica V2 (Rediseñada):** La nueva rúbrica mantuvo la estructura de cuatro componentes pero con criterios más estrictos y penalizaciones claras:
- **R1. Interacción Trazable (20%):** Requisito eliminatorio. Exigía Anexo A completo con transcripción del chat y tiempo verificado ≥ 120 minutos. La ausencia del Anexo A o tiempo insuficiente resultaba en 0 puntos en este ítem y la invalidación de la actividad completa.
- **R2. Estructura del Informe (20%):** Organización clara, etiquetas de módulos M1-M5, tablas y figuras legibles, separación explícita del texto del tutor versus el texto propio del estudiante.
- **R3. Desempeño Técnico (35%):** Corrección de los cálculos de IDR, distribución temporal, distribución areal y SCS-CN; uso correcto de unidades y redondeos; consistencia entre tablas numéricas y narrativa textual.
- **R4. Originalidad y Autoría Técnica (25%):** Similitud textual < 75% en "zonas personales" (verificado por análisis TF-IDF/MinHash/LSH); cinco respuestas de repaso con justificación propia; citación adecuada cuando correspondiera.

El rediseño transformó la evaluación de un modelo centrado en el producto final a uno centrado en el proceso auditable y la producción personal verificable, estableciendo controles que hacían más costoso y menos efectivo el uso instrumental de la IA sin comprensión real.

### 3.5 Arquitectura del Pipeline de Análisis Automatizado

"Para auditar objetivamente las entregas, se diseñó un pipeline técnico reproducible. Primero, se extrajo el texto de los PDF, usando OCR (Tesseract 5.3) cuando fue necesario. Luego de un preprocesamiento estándar (stop-words, lematización con spaCy), se aplicó un análisis dual: vectorización TF-IDF (con scikit-learn) para medir la similitud global del coseno, y firmas MinHash con Locality-Sensitive Hashing (LSH) para detectar similitudes aproximadas (parafraseo). Estos indicadores automatizados, junto con los umbrales operativos (ej. 45% = sospecha), alimentaron la evaluación manual con la rúbrica, permitiendo distinguir copias irregulares de la similitud esperable del texto del tutor.

### 3.6 Recolección de Datos Cualitativos

Además del análisis automatizado, se recopilaron datos cualitativos mediante:
- Análisis de transcripciones de interacción (Anexo A) para estudiantes que lo adjuntaron, evaluando la profundidad del diálogo, la evolución de la complejidad de los prompts y la evidencia de proceso de pensamiento.

- Revisión de respuestas a las cinco preguntas de repaso generadas por el tutor, como indicador de comprensión y apropiación del contenido.
- Documentación de casos especiales (placeholders, irregularidades en transcripciones, texto embebido en imágenes) como evidencia de patrones de mal uso.

Esta triangulación de datos cuantitativos (métricas de similitud, distribuciones de calificaciones) y cualitativos (análisis de chats, evidencias de proceso) permitió una evaluación integral del impacto del rediseño instruccional.

## 4. Resultados

### 4.1 Resultados de la Fase 1: Fracaso Inicial y Hallazgos Críticos

El análisis de la primera entrega (Informe 1 sobre Temas 1-3) reveló un fracaso metodológico generalizado que evidenció las limitaciones críticas de un diseño instruccional sin controles de proceso estrictos.

**Métricas de cumplimiento y evidencia:**

- **Informes con chat exportado:** 1 de 23 (4%)
- **Informes con tiempo verificado ≥ 60 min:** 9 de 23 (39%), aunque la mayoría eran declaraciones genéricas sin marcas temporales auditables
- **Informes con ejercicio numérico propio:** 3 de 23 (13%)
- **Similitud textual ≥ 45%:** 15 de 23 (65%)
- **Copias ≥ 80%:** 2 pares de informes
- **Calificaciones ≥ 6 (aprobados):** 0 de 21 informes válidos (0%)
- **Nota media:** 4.86 / 10

**Hallazgos cualitativos documentados:**

- **Plantillas IA y placeholders:** Frases como "[aquí iría el tiempo]" o "[completar conclusiones]" fueron frecuentes, indicando generación posterior sin sesión guiada real.
- **Falsa documentación del tiempo:** Varios estudiantes declararon bloques genéricos de "1 h" o "2 h" sin sellos horarios. El único chat exportado mostró 17 minutos reales, muy por debajo del mínimo.
- **Ausencia general de ejercicios numéricos:** Apenas 3 informes desarrollaron un cálculo aplicado (Gumbel, Log-Pearson III o IDF). La falta de aplicación práctica contradecía la finalidad del tutor.
- **Copias no declaradas:** Dos pares superaron 80% de similitud léxica. Otros nueve informes compartieron > 55% por uso de la misma plantilla.
- **OCR necesario para PDF con texto embebido como imagen:** Un informe se re-valorizó tras aplicar OCR, aunque igualmente reprobó por falta de evidencia de proceso.

**Conclusión de la Fase 1:** La entrega masiva de informes sin trazabilidad real ni originalidad técnica desembocó en cero aprobaciones cuando se aplicaron criterios de integridad académica estrictos. Sin requisitos explícitos de evidencia de proceso, la mayoría de los estudiantes utilizó la IAGen para generar documentos superficiales, facilitando el plagio y eludiendo el aprendizaje significativo.

## 4.2 Resultados de la Fase 2: Impacto del Rediseño

El análisis de la segunda entrega (Informe 2 sobre Lluvia de Diseño), basada en el Prompt V2 rediseñado y la Rúbrica V2 con controles estrictos, mostró una mejora drástica en la calidad del proceso de aprendizaje y un filtrado efectivo de entregas sin cumplimiento.

**Desempeño global (n=27 entradas de evaluación, incluyendo reelaboraciones):**

- **Media:** 72.9 / 100
- **Desviación estándar:** 26.8
- **Mediana:** 88 / 100 (mejora sustancial respecto a Fase 1)
- **Rango:** 25 - 99
- **Calificaciones ≥ 90/100:** 13 de 27 (48.1%)
- **Calificaciones ≥ 60/100:** 18 de 27 (66.7%)

**Efectividad del control de proceso:**

- **Interacción Trazable (R1) = 0 puntos:** 6 de 27 entradas (22.2%) fueron invalidadas por ausencia del Anexo A o tiempo de interacción insuficiente (< 120 min). Las evaluaciones individuales confirman casos con "No adjuntó el Anexo A" o "Tiempo de interacción < 120 min" como causales de invalidación.
- **Cumplimiento verificado:** Los estudiantes que sí cumplieron con los requisitos (Anexo A completo, ≥ 120 min de interacción verificable, cobertura de módulos M1-M5) alcanzaron informes muy completos y bien estructurados, con altos puntajes en Desempeño Técnico (componente R3) y Estructura (componente R2).

**Distribución por componentes de la rúbrica (promedios):**

- **R1. Interacción (20%):** 11.78 puntos promedio (sobre 20)
- **R2. Estructura (20%):** 14.11 puntos promedio (sobre 20)
- **R3. Desempeño Técnico (35%):** 27.67 puntos promedio (sobre 35)
- **R4. Originalidad (25%):** 19.33 puntos promedio (sobre 25)

**Calidad técnica:** Los estudiantes con cumplimiento completo demostraron: resolución paso a paso del ejemplo numérico estándar (cuenca de 10 km²), correcta aplicación de métodos IDR, distribución temporal (SCS II/III), distribución areal, método SCS-CN para abstracciones, integración al hietograma de diseño y cálculo de precipitación efectiva. Los mejores casos incluyeron justificación de parámetros, verificaciones numéricas y pequeñas adaptaciones (ej. análisis con cuencas de 75 km² aplicando criterio ingenieril).

**Casos ejemplares:** Estudiantes con calificaciones de 9.9 y 9.7 presentaron chats completos de 150-210 minutos, cobertura completa de M1-M5, ejemplo numérico resuelto con unidades, respuestas propias a las cinco preguntas de repaso con justificación técnica, y evidencia clara de proceso de aprendizaje genuino.

## 4.3 Análisis de Originalidad y Autoría Técnica (Fase 2)

El pipeline automatizado TF-IDF/LSH permitió un análisis matizado de la originalidad y la detección focalizada de irregularidades académicas.

**Bloques doctrinales previsibles:** El análisis TF-IDF confirmó que las explicaciones teóricas estándar (ej. método SCS-CN, conceptos de IDR local, distribuciones de probabilidad) presentaban alta similitud entre estudiantes. Esto era esperable, ya que provenían del propio tutor y del material base cargado, y no implicaba plagio. Este hallazgo validó la decisión metodológica de no sancionar similitud automáticamente, sino de interpretar la métrica en función del tipo de contenido.

**"Zonas personales auditables":** La mayor variabilidad y donde se exigía originalidad se encontró en: (a) la resolución del ejemplo numérico con la cuenca de 10 km² (o variaciones), incluyendo cálculos intermedios con unidades, justificación de coeficientes (ej. número de curva CN), y presentación de resultados; y (b) las respuestas a las "5 preguntas de repaso" personalizadas generadas por el tutor al final de la sesión. En estas zonas, MinHash/LSH fue altamente efectivo para distinguir la reescritura auténtica (estudiante que explica con sus palabras) del copiado entre pares (fragmentos idénticos o casi idénticos sin fuente común).

**Detección de irregularidades mediante análisis de chats (Anexo A):** El análisis cualitativo de las transcripciones completas (cuando estaban presentes) fue crucial para detectar irregularidades sutiles. Por ejemplo, en el caso del alumno "Lucena Ignacio", la revisión detallada del chat reveló que el Módulo 5 de su transcripción contenía un ejercicio con el nombre y datos de otro alumno ("Marcelo Exequiel Rivadeneira"), evidenciando que había compartido o copiado parte de la sesión de otro estudiante. Esta irregularidad grave solo pudo detectarse gracias a la exigencia del Anexo A completo con trazabilidad, y no habría sido visible analizando únicamente el informe final.

**Conclusión del análisis de similitud:** El enfoque combinado TF-IDF (para similitud global) + MinHash/LSH (para similitud aproximada robusta a parafraseo) + revisión humana focalizada en zonas personales permitió separar eficazmente: (i) la similitud estructural esperable (texto canónico del tutor presente en todos los informes) de (ii) las copias irregulares entre estudiantes, y (iii) la falta de producción personal en secciones donde se exigía originalidad. Este análisis técnico funcionó como "semáforo" para priorizar la revisión docente, no como herramienta de sanción automática.

## 4.4 Comparación Fase 1 vs. Fase 2: Evidencia del Impacto del Rediseño

La comparación cuantitativa entre ambas fases evidencia de manera contundente el impacto del rediseño instruccional:

- **Tasa de aprobación:** Fase 1: 0% | Fase 2: 66.7% (con criterios igualmente estrictos)
- **Nota media:** Fase 1: 4.86 / 10 | Fase 2: 72.9 / 100 (escala homologada: ≈ 7.3 / 10)
- **Mediana:** Fase 1: No reportada (distribución muy baja) | Fase 2: 88 / 100
- **Evidencia de proceso (chat exportado):** Fase 1: 4% | Fase 2: Obligatorio (invalidación si ausente)

- **Ejercicio numérico verificable:** Fase 1: 13% | Fase 2: Obligatorio en módulo M5

**Interpretación:** El rediseño no "facilitó" la aprobación relajando criterios, sino que transformó la naturaleza de la tarea de manera que el cumplimiento genuino (interacción real, proceso de pensamiento documentado, producción técnica verificable) fuera la vía más directa para obtener una buena calificación. Los estudiantes que intentaron eludir el proceso mediante uso instrumental de la IA fueron efectivamente filtrados por los requisitos eliminatorios (Anexo A, tiempo mínimo). Los que se comprometieron con el proceso guiado alcanzaron altos niveles de desempeño técnico y comprensión.

## 5. Discusión

### 5.1 Aprendizaje Significativo vs. Uso Superficial de la IA

El patrón empírico documentado en este estudio es consistente y claro: en ausencia de controles de proceso explícitos, la IAGen enmascara efectivamente la falta de apropiación conceptual por parte del estudiante. La Fase 1 demostró que el estudiante, motivado por la búsqueda de eficiencia ("ganancias de productividad horaria") documentada en la literatura argentina, puede entregar un producto final que aparenta corrección formal pero que carece completamente de rastro de comprensión profunda o proceso de construcción del conocimiento.

Este hallazgo confirma la tensión fundamental identificada en el marco teórico: la divergencia entre la búsqueda de eficiencia del estudiante y la búsqueda de profundidad del educador no puede resolverse mediante la mera introducción de la tecnología, sino que exige un rediseño del andamiaje pedagógico y evaluativo que alinee los incentivos.

La Fase 2 demuestra que el control del proceso invierte esta dinámica de manera efectiva. Cuando se fuerza la explicitación del razonamiento mediante mecanismos auditables (Anexo A con transcripción completa y marcas temporales, tiempo mínimo verificado de 120 minutos, resolución de ejercicio numérico con cálculos intermedios documentados, respuestas propias a preguntas de repaso personalizadas), emergen entregas con alta calidad técnica y evidencia clara de apropiación conceptual. El rediseño hizo que el "camino fácil" (uso instrumental sin comprensión) fuera estructuralmente menos viable que el "camino correcto" (compromiso genuino con el proceso guiado).

La conclusión metodológica central es que la evaluación en la era de la IAGen debe migrar decididamente de la evaluación del "producto" (el informe final, que la IA puede generar eficientemente) a la evaluación del "proceso" (la trayectoria de construcción del conocimiento, que solo el estudiante puede realizar). Este cambio no es opcional sino imperativo si se busca preservar la integridad académica y el aprendizaje significativo.

### 5.2 Integridad Académica y el Rol del Análisis de Similitud

El hallazgo más decisivo de este estudio en términos de integridad académica es que el control de proceso mediante evidencia auditable (Anexo A con transcripción completa, metadatos temporales, checksums opcionales) es un mecanismo de verificación significativamente más efectivo y robusto que cualquier detector genérico de texto generado por IA.

Los detectores de IA basados en patrones probabilísticos del lenguaje (como GPTZero, ZeroGPT, o los módulos de detección de Turnitin) han demostrado tener tasas de falsos positivos significativas y ser vulnerables a estrategias de elusión mediante parafraseo o instrucciones al modelo generativo ("reescribe este texto para que no sea detectado"). Más fundamentalmente, estos detectores operan en el paradigma equivocado: intentan responder "¿fue escrito por IA?" cuando la pregunta relevante pedagógicamente es "¿hubo aprendizaje?"

El enfoque de este estudio redefine el rol de las herramientas técnicas de análisis textual (TF-IDF, MinHash/LSH). En lugar de ser instrumentos punitivos de "detección de plagio", se convierten en técnicas auxiliares de diagnóstico no punitivo. Su función es ayudar al docente a focalizar la revisión humana donde debe haber originalidad — en las "zonas personales auditables" (síntesis con palabras propias, justificación técnica de decisiones, cálculos con unidades)—, separando la similitud esperable (texto canónico del tutor, explicaciones teóricas estándar) de la copia irregular entre estudiantes.

La política de umbrales operativos implementada (30% ruido; 45% media; 75% alta) no se utilizó como criterio automático de sanción, sino como "semáforo" para priorizar atención docente. Un informe con 65% de similitud no es automáticamente inválido; el análisis cualitativo determina si esa similitud proviene de secciones donde era esperable (explicaciones del tutor) o de zonas donde se exigía producción personal.

Este enfoque tiene una ventaja adicional: es robusto ante la evolución tecnológica. Independientemente de cuán sofisticados se vuelvan los modelos de IAGen, la exigencia de evidencia de proceso (chat exportado, marcas temporales, preguntas de repaso respondidas en vivo) permanece como control efectivo, porque la IA no puede retroactivamente "crear" una sesión de 120 minutos de interacción genuina.

### 5.3 Implicancias para la Ingeniería y la Educación Superior en América Latina

Este estudio responde directamente a la convocatoria del dossier temático de ESS sobre el contexto específico de América Latina. En la región, y particularmente en universidades públicas como la UTN-FRT, los sistemas educativos enfrentan desafíos estructurales que hacen especialmente relevante (y compleja) la integración de tecnologías como la IAGen.

En asignaturas técnicas con fuerte componente de cálculo aplicado (como Hidrología y Obras Hidráulicas, pero extensible a Estructuras, Geotecnia, Hidráulica de Canales, y otras en carreras de ingeniería), los tutores de IA bien diseñados pueden cumplir un rol de "nivelación" de acceso a retroalimentación técnica. En contextos donde la masificación estudiantil y los cuellos de botella docentes limitan la capacidad de atención personalizada, un tutor conversacional que puede atender 24/7, en español, con paciencia infinita y adaptándose al ritmo del estudiante, representa una oportunidad genuina de democratización del acceso a apoyo académico de calidad.

Sin embargo, el estudio también demuestra que esta promesa solo se materializa si el diseño instruccional: (i) exige evidencia auditable del proceso de aprendizaje, no solo del producto final; (ii) separa explícitamente el texto generado por la IA del texto que debe producir el estudiante, estableciendo "zonas personales auditables"; y (iii) evalúa la producción técnica verificable (cálculos con unidades, justificación de parámetros, coherencia entre narrativa y resultados numéricos).

El protocolo de cinco componentes propuesto (prompts modulares con checkpoints, trazabilidad obligatoria vía Anexo A y tiempo mínimo, zonas personales auditables, rúbrica de proceso, analítica de similitud localizada) es escalable y transferible a otras asignaturas STEM en la región. Su implementación requiere inversión inicial en diseño instruccional (redacción de prompts, definición de rúbricas, configuración del pipeline técnico) pero, una vez establecido, el sistema es sostenible y puede gestionarse con los recursos docentes habituales.

Una consideración crítica para el contexto regional es la equidad en el acceso a la tecnología. Si bien las versiones gratuitas de ChatGPT y otras IAGen están ampliamente disponibles, existen brechas en conectividad, alfabetización digital y familiaridad con estas herramientas. La implementación de tutores de IA debe acompañarse de sesiones de capacitación (onboarding) obligatorias que aseguren que todos los estudiantes, independientemente de su experiencia previa, puedan interactuar efectivamente con la herramienta. El estudio de caso documenta que la "pedagogía del prompt" —enseñar explícitamente a formular preguntas complejas y estructuradas— es una competencia que se desarrolla, no una habilidad innata.

### 5.4 Limitaciones del Estudio

Este estudio presenta varias limitaciones que deben reconocerse al interpretar sus hallazgos y considerar su generalización:

- **Tamaño de muestra limitado:** El estudio se basa en una única cohorte de 23 estudiantes en una sola asignatura de una única institución. Si bien el diseño de estudio de caso profundo es apropiado para exploración metodológica, la generalización estadística a otras poblaciones, asignaturas o contextos institucionales debe hacerse con cautela.

- **Ausencia de grupo control:** El estudio no incluye un grupo control contemporáneo (estudiantes de la misma cohorte que no utilizaran el tutor de IA), lo que limita la capacidad de atribuir causalmente las mejoras observadas exclusivamente a la intervención. La comparación con cohortes históricas anteriores proporciona evidencia contextualizadora pero no experimental.

- **Especificidad disciplinar:** La asignatura Hidrología y Obras Hidráulicas tiene características particulares (fuerte componente de cálculo numérico, existencia de métodos estandarizados, necesidad de justificación de parámetros) que hacen especialmente viable el enfoque de "zonas personales auditables" con ejercicios numéricos. La transferibilidad a asignaturas de humanidades, ciencias sociales o incluso otras ramas de la ingeniería con menor componente cuantitativo requeriría adaptaciones del protocolo.

- **Dependencia de una plataforma específica:** El tutor se implementó sobre ChatGPT (GPT-4) durante 2024-2025. La evolución tecnológica acelerada en este campo implica que las capacidades, limitaciones y comportamientos de la herramienta pueden cambiar. Estudios de replicación con otras plataformas de IAGen (Claude, Gemini, LLaMA) o con versiones futuras de GPT serían valiosos.

- **Carga docente incrementada:** El rediseño implicó una inversión significativa de tiempo docente en el análisis de chats exportados (Anexo A), revisión de cálculos intermedios y evaluación cualitativa de respuestas a preguntas de repaso. Si bien el pipeline automatizado de similitud reduce parcialmente esta carga, el modelo no es tan "escalable" como la evaluación tradicional basada únicamente en producto final. La viabilidad en cursos masivos (>100 estudiantes) requeriría investigación adicional y posible soporte de asistentes de enseñanza.

A pesar de estas limitaciones, el estudio aporta evidencia empírica valiosa sobre un fenómeno poco documentado en la región y propone un protocolo metodológico replicable que puede servir como punto de partida para investigaciones futuras más amplias.

## 6. Conclusiones y Propuestas Replicables

Este estudio de caso demuestra empíricamente que la innovación pedagógica con Inteligencia Artificial Generativa en educación en ingeniería rinde frutos significativos solo cuando el diseño instruccional controla rigurosamente el proceso de aprendizaje y evalúa la producción personal auditable, no únicamente el producto final. La transición documentada de la Fase 1 (fracaso generalizado con 0% de aprobación) a la Fase 2 (mediana de 88/100 con 66.7% de aprobación) no fue resultado de una innovación tecnológica, sino de un rediseño pedagógico que alineó incentivos, estableció controles de proceso y transformó la naturaleza de la evaluación.

### 6.1 Contribuciones Principales

El estudio realiza tres contribuciones principales al campo de la innovación educativa con IAGen:

- **Evidencia empírica desde un contexto subrepresentado:** Documenta de manera exhaustiva una intervención completa (diseño, fracaso, rediseño, éxito) en una universidad pública latinoamericana, contribuyendo a llenar el vacío de estudios de caso empíricos en la región identificado en la introducción.

- **Protocolo metodológico transferible:** Propone un marco de cinco componentes (descrito a continuación) que es replicable en otras asignaturas STEM, escalable a diferentes tamaños de cohorte y adaptable a distintos contextos institucionales.

- **Pipeline técnico reproducible:** Detalla una arquitectura de análisis automatizado (OCR, TF-IDF, MinHash/LSH) que permite auditar objetivamente la similitud textual y priorizar la revisión docente sin depender de detectores genéricos de IA, que son poco confiables.

### 6.2 Prospectiva: Hacia una Educación Superior Aumentada e Íntegra

La educación superior debe evolucionar desde un modelo centrado en la transmisión y evaluación de información —un modelo que la IA ha vuelto estructuralmente obsoleto, dado que cualquier estudiante puede ahora acceder instantáneamente a explicaciones técnicas correctas y textos bien estructurados— hacia un modelo enfocado en el desarrollo de competencias metacognitivas, creativas y críticas que permanecen exclusivamente humanas.

El futuro de la formación de profesionales, como los ingenieros de este estudio, dependerá crucialmente de su capacidad para: (a) colaborar inteligentemente con sistemas de IA, reconociendo sus fortalezas (velocidad, escala, acceso a patrones en grandes corpus) y sus limitaciones (falta de comprensión genuina, ausencia de intuición física, incapacidad de juicio ético contextualizado); (b) cuestionar, validar y contextualizar críticamente la información generada por algoritmos, no aceptándola ciegamente; (c) aplicar juicio profesional fundado en principios éticos, conocimiento de contexto local y consideración de impactos sociales y ambientales a la resolución de problemas complejos que definen nuestro tiempo.

Estas competencias no se desarrollan mediante exposición pasiva a la tecnología ni mediante prohibiciones, sino mediante el diseño intencional de experiencias de aprendizaje que posicionen a la IA como andamiaje para el pensamiento de orden superior, no como sustituto de él. El protocolo propuesto en este artículo representa una contribución concreta y operacionalizable hacia ese objetivo.

Para instituciones de educación superior en la región que estén considerando la integración de IAGen, las lecciones aprendidas de este estudio de caso pueden resumirse en un imperativo: no basta con habilitar el acceso a la herramienta; es imprescindible rediseñar simultáneamente las tareas, los criterios de evaluación y los mecanismos de evidencia del proceso de aprendizaje. Sin ese rediseño sistémico, la IAGen facilitará el atajo y la superficialidad, no el aprendizaje profundo.

Finalmente, este estudio nos confronta con una verdad incómoda pero necesaria: no existe práctica pedagógica perfecta, ni diseño instruccional infalible, ni tecnología —por avanzada que sea— capaz de sustituir el elemento más fundamental y esquivo del aprendizaje: la voluntad genuina del estudiante de aprender. Las herramientas, los tutores de IA, las rúbricas exigentes y los controles de proceso pueden crear las condiciones propicias, pueden eliminar obstáculos, pueden hacer visible lo que antes permanecía oculto, pueden incluso desincentivar el fraude. Pero ninguna de ellas puede encender la chispa de la curiosidad intelectual en quien solo busca cumplir, aprobar, avanzar sin transformarse. El fracaso de la Fase 1 y el éxito de la Fase 2 no se explican únicamente por la calidad del diseño —aunque esta fue determinante—, sino también por la disposición de algunos estudiantes a comprometerse con un proceso de aprendizaje genuino cuando se les presentó una estructura que lo hacía posible y valorable. Un profesor, por dedicado y experto que sea, no puede guiar el aprendizaje de quien ha decidido no aprender; solo puede, con honestidad y rigor, mantener abierta la puerta y señalar el camino para quienes elijan atravesarla. La integración de la Inteligencia Artificial en la educación superior no altera esta verdad ancestral; simplemente la hace más evidente, más urgente de reconocer y, quizás, más difícil de eludir.

## Referencias